\documentclass[twocolumn,amsmath,prb,letterpaper]{revtex4} 
\usepackage{graphicx}

\newcommand{\bt}{\begin{tabbing}}
\newcommand{\et}{\end{tabbing}}

\newcommand{\ignore}[1]{}

\def\gtwid{\mathrel{\raise.3ex\hbox{$>$\kern-.75em\lower1ex\hbox{$\sim$}}}}
\def\ltwid{\mathrel{\raise.3ex\hbox{$<$\kern-.75em\lower1ex\hbox{$\sim$}}}}
\def\pibox{\mathrel{\raise.3ex\hbox{$\pi$\kern-.75em\lower2ex\hbox{$\Box$}}}}

\def\abs{{\scriptscriptstyle ||}}

\begin{document}

\title{Canonical and Grand Canonical Ensemble Expectation Values\\ from Quantum Monte Carlo Simulations}

\author{R.D.~Sedgewick, D.J.~Scalapino and R.L.~Sugar}
\affiliation{Department of Physics\\ 
University of California, Santa Barbara, CA 93106}

\author{L.~Capriotti}
\affiliation{Kavli Institute for Theoretical Physics\\ 
University of California, Santa Barbara, CA 93106}


\begin{abstract}

We show how canonical ensemble expectation values can be extracted
from quantum Monte Carlo simulations in the grand canonical ensemble.
In order to obtain results for all particle sectors, a modest number
of grand canonical simulations must be performed, each at a different
chemical potential.  From the canonical ensemble results, grand
canonical expectation values can be extracted as a continuous function
of the chemical potential.  Results are presented from the application
of this method to the two-dimensional Hubbard model.

\end{abstract}
\pacs{71.10.Fd, 71.15.Dx, 05.10.Ln}
\maketitle

\section{Introduction}

The properties of strongly correlated electron systems near the Mott
insulating phase depend sensitively on the doping.\cite{REVIEW}  Most
simulations of these systems have been carried out within the grand
canonical ensemble, where a convenient formalism exists for the
evaluations of finite temperature Green's functions and other physical
quantities that can be expressed in terms of them.\cite{DET} In this
framework, to explore the doping dependence, it is necessary to carry
out simulations at various discrete values of the chemical potential
$\{\mu_\alpha\}$, and interpolate between them. In this paper we
describe a method for optimally combining data from these simulations
to obtain results for a continuous range of $\mu$.  We first evaluate
canonical ensemble expectation values and partition functions (up to
an overall constant) from simulations performed in the grand canonical
ensemble. Using the approach of Ferrenberg and Swendsen~\cite{FS} to
combine results from simulations performed with different value of the
chemical potential, we obtain canonical ensemble quantities for a wide
range of fillings from a modest number of simulations.  From these we
are able to construct grand canonical expectation values, enabling us
to study a variety of physical quantities as a continuous function of
the chemical potential.

In Section~II we present our methodology, and in Section~III we
illustrate it with results for the two-dimensional Hubbard model.

\section{Methodology}

We begin by briefly summarizing the approach to the simulation of
strongly correlated many-electron systems in the grand canonical ensemble
set out in Ref.~\onlinecite{DET}.
The expectation value of a physical observable $O$ is
\begin{equation}
\langle O(\mu)\rangle = \frac{{\rm Tr}\, \left[O\,e^{-\beta (H -\mu N)}\right]}
         {{\rm Tr}\,\left[e^{-\beta (H -\mu N)}\right]},
\label{eq:trace}
\end{equation}
where $H$ is the Hamiltonian, $\beta$ the inverse temperature, $\mu$ the
chemical potential and $N$ the number operator for the electrons.
In order to perform a numerical simulation, one must first evaluate
the traces over the electron degrees of freedom. This is possible
if the Hamiltonian is quadratic in the electron creation and annihilation
operators, or can be made so through a Hubbard-Stratonovich transformation.
To this end we introduce a small imaginary-time step, $\Delta\tau$, by 
writing $\beta=\Delta\tau\, L$. The partition function can then be
written in the form
\begin{equation}
Z(\mu)={\rm Tr}\, \left[e^{-\beta (H -\mu N)}\right] = 
{\rm Tr}\,\left[e^{-\Delta\tau(H-\mu N)}\right]^L.
\end{equation}
For each time-slice $\ell=1\ldots L$, we introduce a set of
Hubbard-Stratonovich variables $x(\ell)$ such that
\begin{equation}
e^{-\Delta\tau\, H} = \sum_{x(\ell)}\; \omega\left(x(\ell)\right)\; 
e^{-\Delta\tau\; \sum_{i,j,\sigma}\; c^\dagger_{i\,\sigma}\;
h^\sigma_{i,j}\left(x(\ell)\right) \; c^{}_{j\,\sigma}}.
\label{eq:hs}
\end{equation}
Here $c^\dagger_{i\,\sigma}$ and $c^{}_{i\,\sigma}$ are the creation
and annihilation operators for electrons at lattice site $i$ with
$z$-component of spin $\sigma$; $h_{i, j}^\sigma\left(x(\ell)\right)$
is a single particle Hamiltonian for an electron propagating through
the external field, $x(\ell)$; and $\omega\left(x(\ell)\right)$ is a
positive definite weight function.  $\sum_{x(\ell)}$ indicates an
integral over continuous Hubbard-Stratonovich variables or a sum over
discrete ones. Typically, $x(\ell)$ has a component for each spatial
lattice site or link.

The traces in Eq.~(\ref{eq:trace}) can now be performed, yielding an expression
of the general form
\begin{equation}
\langle O(\mu)\rangle = \frac{\sum_x\: \rho(x,\mu)\, O(x,\mu)}
{\sum_{x^\prime}\: \rho(x^\prime,\mu)}.
\label{eq:sum}
\end{equation}
Here $x$ stands for the totality of Hubbard-Stratonovich variables on all
time slices, 
\begin{equation}
\rho(x,\mu) = D_\uparrow(x,\mu)\; D_\downarrow(x,\mu)\; \prod_{\ell=1}^L\; \omega\left(x(\ell)\right)
\end{equation}
and the determinants for spin up and down electrons are given by
\begin{equation}
D_\sigma(x,\mu) = {\rm Det}\: \left[I +\, e^{\,\beta\,\mu}\: A_\sigma(x)\right],
\label{eq:det}
\end{equation}
where $I$ is the unit matrix and
\begin{equation}
A_\sigma(x) = e^{-\Delta\tau h^\sigma(x(L))}\ldots e^{-\Delta\tau h^\sigma(x(1))}.
\label{eq:asig}
\end{equation}
The quantity $O(x,\mu)$ in Eq.~(\ref{eq:sum}) can generally be
expressed in terms of finite temperature Green's functions for a
single electron propagating in the background field provided by the
Hubbard-Stratonovich variables, $x$.  For example,
\begin{eqnarray}		
{\rm Tr}\; \left[c^{}_{i\,\sigma}\;c^\dagger_{j\,\sigma^\prime}\; e^{-\beta (H -\mu N)}\right]
= & D_\uparrow(x,\mu)\; D_\downarrow(x,\mu)\; \delta_{\sigma,\sigma^\prime}
\nonumber \\
& \times \left(\frac{1}{I+e^{\,\beta\,\mu}\:
A_\sigma(x)}\right)_{i,j}.
\end{eqnarray}
For models with particle-hole symmetry, such as the Hubbard model at 
half-filling,  the product of the electron determinants is positive, 
and one can use importance sampling techniques to generate a sequence 
of Hubbard-Stratonovich configurations with the probability distribution
\begin{equation}
P(x,\mu) = \frac{\rho(x,\mu)}{\sum_{x^\prime}\: \rho(x^\prime,\mu)}.
\end{equation}
The average value of $O(x,\mu)$ in these configurations is then an
estimator for $\langle O(\mu)\rangle$. Details of an algorithm for efficiently 
generating configurations are given in Ref.~\onlinecite{DET}.

For systems which do not have particle-hole symmetry, such as the
Hubbard model away from half-filling, the product of electron determinants
will in general not be positive definite. In such cases, one can generate
Hubbard-Stratonovich fields using the probability distribution
\begin{equation}
P_{\abs}(x,\mu) = \frac{|\rho(x,\mu)|}{\sum_{x^\prime}\: |\rho(x^\prime,\mu)|}.
\end{equation}
It is then necessary to move the sign of $\rho(x,\mu)$,
\begin{equation}
S(x,\mu)=\frac{\rho(x,\mu)}{|\rho(x,\mu)|}=\pm 1
\end{equation}
into the measurements yielding
\begin{eqnarray}
\langle O(\mu)\rangle &=& \sum_x\; P(x,\mu)\: O(x,\mu) \nonumber \\
&=& \frac{\sum_x \; 
P_{\abs}(x,\mu)\; O(x,\mu)\: S(x,\mu)}
{\sum_{x^\prime}\; P_{\abs}(x,\mu)\;S(x,\mu)}.
\end{eqnarray}
The expectation value of the sign can be written
\begin{eqnarray}
\langle S(\mu)\rangle &=& \sum_x\; 
P_{\abs}(x,\mu)\; S(x,\mu) \nonumber \\ 
&=& 
\frac{\sum_x\; \rho(x,\mu)}{\sum_{x^\prime}\; |\rho(x^\prime,\mu)|}
=\frac{Z(\mu)}{Z_{\abs}(\mu)}.
\end{eqnarray}
Here $Z(\mu)$ is the partition function of the physical system,
and $Z_\abs(\mu)$ that of a fictitious one in which the sign
of the product of determinants, $S(x,\mu)$, is ignored.

To obtain information about the canonical ensemble from grand canonical
simulations, we note that so long as the electron number operator commutes
with the Hamiltonian, and the Hubbard-Stratonovich variables are chosen 
so that Eq.~(\ref{eq:hs}) holds, then the product of electron determinants 
has an expansion in the fugacity of the form
\begin{equation}
D_\uparrow(x,\mu)\: D_\downarrow(x,\mu)= \sum_N\; Z_N(x)\; e^{\,\beta\,\mu\, N}.
\label{eq:fug}
\end{equation}
Once we have gone to the computational expense of perform an LDU
decomposition of $A_\sigma(x)$, Eq.~(\ref{eq:asig}), which we must do
each time we make a measurement,\cite{DET} it is straightforward to
evaluate the left-hand side of Eq.~(\ref{eq:fug}) for a number of
different values of $\mu$. Eq.~(\ref{eq:fug}) then yields a set of
linear equations that can be solved for the $Z_N(x)$. At moderate to
low temperatures, only a limited subset of the $Z_N(x)$ will make a
significant contribution to the product of determinants, so the system
of equations to be solved is considerably smaller than the number of
spatial lattice points. Since the canonical partition function for the
sector with electron number $N$ is given by
\begin{equation}
\sum_x\; \frac{P_\abs(x,\mu)\;Z_N(x)}{|D_\uparrow(x,\mu)\:D_\downarrow(x,\mu)|}
= \frac{Z_N}{Z_\abs(\mu)} 
\equiv {\widetilde Z }_N(\mu),
\label{eq:zn}
\end{equation}
where
\begin{equation}
Z_N=\sum_x\; Z_N(x) \prod_{\ell=1}^L\; \omega\left(x(\ell)\right),
\end{equation}
we can evaluate ${\widetilde Z}_N$ using an ensemble
of Hubbard-Stratonovich fields generated with the probability
distribution $P_\abs(x,\mu)$.

If the operator $O$ is defined on a single imaginary-time slice, or if it
does not change the electron number from time slice to time slice, then
we can also write
\begin{equation}
O(x,\mu)\; D_\uparrow(x,\mu)\:D_\downarrow(x,\mu)= \sum_N\; O_N(x)\; e^{\,\beta\,\mu\, N},
\label{eq:ox}
\end{equation}
and we can obtain a set of linear equations for the $O_N(x)$ by evaluating 
the left-hand side of Eq.~(\ref{eq:ox}) for different values of $\mu$.
In this case
\begin{equation}
  \sum_x\; \frac{P_\abs(x,\mu)\; O_N(x)}{|D_\uparrow(x,\mu)\: D_\downarrow(x,\mu)|} = 
  \frac{O_N}{Z_\abs(\mu)} 
  \equiv {\widetilde O}_N(\mu).
  \label{eq:on}
\end{equation}
Finally, the expectation value of the operator $O$ in the canonical ensemble
sector with electron number $N$ is
\begin{equation}
  \langle O\rangle_N = \frac{\widetilde O_N}{{\widetilde Z}_N}=\frac{O_N}{Z_N}.
\end{equation}
Also, once the $Z_N$ and $O_N$ are in hand, 
\begin{equation}
\langle O \rangle = \frac{\sum_N O_N e^{\beta\mu N}}{\sum_N Z_N
  e^{\beta \mu N}}
\label{eq:Oexpect}
\end{equation}
gives the grand canonical
expectation values as continuous functions of $\mu$.


From simulations at a single value of $\mu$ one only expects to
be able to make accurate determinations of the $Z_N$ and $O_N$ for $N$
in the vicinity of $\langle N\rangle$. We must therefore perform a set
of simulations with chemical potentials $\mu_\alpha$, sufficiently
spaced to cover the range of $N$ relevant to the problem of
interest. As indicated in Eqs.~(\ref{eq:zn}) and (\ref{eq:on}), the
outputs of our simulations are ${\widetilde
Z}_N(\mu_\alpha)=Z_N/Z_\abs(\mu_\alpha)$ and ${\widetilde
O}_N(\mu_\alpha)=O_N/Z_\abs(\mu_\alpha)$, rather than $Z_N$ and $O_N$.
We can combine results from simulations with different values of the
chemical potential by writing
\begin{eqnarray}
Z_N & = & \sum_\alpha \; c_\alpha \; {\widetilde Z}_N(\mu_\alpha) \; Z_\abs(\mu_\alpha)\\
O_N & = & \sum_\alpha \; d_\alpha \; {\widetilde O}_N(\mu_\alpha) \; Z_\abs(\mu_\alpha),
\label{eq:combo}
\end{eqnarray}
with
\begin{equation}
\sum_\alpha \; c_\alpha = \sum_\alpha \; d_\alpha = 1.
\label{eq:constraint}
\end{equation}
Following Ferrenberg and Swendsen, we choose the $c_\alpha$ and $d_\alpha$
to minimize the variance of $Z_N$ and $O_N$ subject to the constraints
of Eq.~(\ref{eq:constraint}). A short calculation yields
\begin{equation}
c_\alpha = \frac{1/[Z_\abs(\mu_\alpha)\; \sigma^2_N(\mu_\alpha)]}
{\sum_\gamma \; 1/[Z_\abs(\mu_\gamma)\; \sigma^2_N(\mu_\gamma)]},
\label{eq:solve}
\end{equation}
where $\sigma^2_N(\mu_\alpha)$ is the variance of ${\widetilde
Z}_N(\mu_\alpha)$, which we determine from the simulation. Of course,
a corresponding result holds for the $d_\alpha$ with
$\sigma^2_N(\mu_\alpha)$ replaced by the variance of the ${\widetilde
O}_N(\mu_\alpha)$.

The constants $Z_\abs(\mu_\alpha)$ can be determined up to an overall 
normalization by iteratively solving the equation
\begin{equation}
Z(\mu_\alpha) = Z_\abs(\mu_\alpha)\; \langle S(\mu_\alpha)\rangle
= \sum_N \; Z_N \; e^{\,\beta\,\mu_\alpha\, N}
\label{eq:z}
\end{equation}
with the $Z_N$ given by Eqs.~(\ref{eq:combo}) and (\ref{eq:solve}). The
$\langle S(\mu_\alpha)\rangle$ are measured directly in the
simulations. 

It is also possible to obtain $Z_\abs(\mu)$, and therefore
$\langle S(\mu)\rangle$, as a continuous function of $\mu$.
We simply note that 
\begin{equation}
  Z_\abs(\mu) = \sum_N \; {Z_N}_\abs(\mu)\; e^{\,\beta\,\mu\, N},
\label{eq:zp}
\end{equation}
where
\begin{equation}
{Z_N}_\abs(\mu) = \sum_x \; Z_N(x)\; S(x,\mu) \prod_{\ell=1}^L\;
\omega\left(x(\ell)\right).
\label{eq:zNp}
\end{equation}

Note that once the $Z_N(x)$ are known for any configuration, we can
determine $S(x,\mu)$ for any value of $\mu$ from the right-hand side
of Eq.~(\ref{eq:fug}).  A simulation performed at the chemical
potential, $\mu_\alpha$, will, of course, only determine the ratio
${Z_N}_\abs(\mu)/Z_\abs(\mu_\alpha)$. However, we can combine results
from simulations performed at different values of the chemical
potential just as for the $Z_N$.

\section{Numerical Results}

\begin{figure}
\centerline{\includegraphics[width=3.5in]{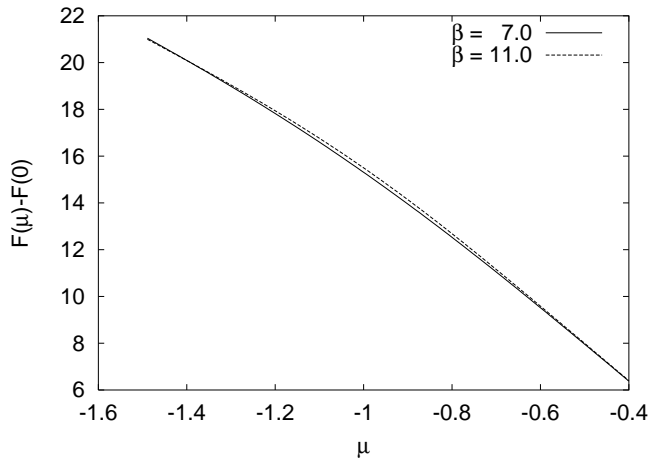}}
\caption{\label{fig:F} Free energy difference between half-filled
  system and system with chemical potential $\mu$.  Statistical errors
  are negligible on this scale.}
\end{figure}

We illustrate the methodology outlined in the last section with
results for the two-dimensional Hubbard model. The Hamiltonian is
\begin{equation}
H= -t \sum_{<ij>,\sigma}\; c^\dagger_{i\,\sigma}\, c^{}_{j\,\sigma} 
+ c^\dagger_{j\,\sigma}\, c^{}_{i\,\sigma}
+ U\, \sum_i\; (n_{i\uparrow}-{\scriptstyle \frac{1}{2}})(n_{i\downarrow}-{\scriptstyle \frac{1}{2}}).
\end{equation}
Here $c^\dagger_{i\,\sigma}$ and $c^{}_{i\,\sigma}$ are the creation and
annihilation operators for electrons with $z$-component of spin
$\sigma$ at lattice site $i$, and
$n_{i\,\sigma}=c^\dagger_{i\,\sigma}\,c^{}_{i\,\sigma}$.  The sum $<ij>$
is over all pairs of nearest neighbor lattice sites.  $t$ is the
hopping parameter, and $U$ the Coulomb coupling constant.


The Coulomb term is reduced to quadratic form in the electron creation
and annihilation operators using Hirsch's discrete Hubbard-Stratonovich 
transformation~\cite{HIRSCH}
\begin{multline}
e^{-\Delta\tau\: U\,(n_{i\uparrow}-\frac{1}{2})(n_{i\downarrow}-\frac{1}{2})}
= \\
 {\scriptstyle\frac{1}{2}}\: e^{-\Delta\tau\: U/4} \! 
 \sum_{x_i(\ell)=\pm 1}
e^{-\Delta\tau\: x_i(\ell)\:\lambda\:(n_{i\uparrow}-n_{i\downarrow})}.
\end{multline}
For $\mu=0$, which corresponds to half-filling, particle-hole symmetry
implies that $D_\uparrow(x,0)\,D_\downarrow(x,0)$ is always
positive,\cite{HIRSCH} so $S(x, 0)=1$, and there is no sign problem.
It is therefore convenient to adopt the normalization
$Z_\abs(0)=Z(0)=1$ in solving Eq.~(\ref{eq:z}) for
$Z_\abs(\mu_\alpha)$. Thus, we are in fact able to use Eqs.~(\ref{eq:z})
and (\ref{eq:zp}) to determine $Z(\mu)/Z(0)$ and $Z_\abs(\mu)/Z_\abs(0)$
respectively.

All of the results we present here are on a $4\times 4$ lattice with
$t=1$ and $U=4$.  The number of time slices, $L$, is chosen so that
$\Delta \tau = 1/8$.  Except where otherwise noted, simulations
were performed at $\mu=-1.025$ and $\mu=-0.6$ for each temperature.
At $\beta=8$ we performed additional runs at both $\mu=-0.9625$ and
$\mu=-0.9$, while at other temperatures either $\mu=-0.9625$ or
$\mu=-0.90$ was used.  For runs at $\mu=-0.6$, 100,000 Monte Carlo
sweeps with 10,000 warmup sweeps were performed.  For all other runs,
400,000 Monte Carlo sweeps with 10,000 warmup sweeps were performed.
For all simulations, non-local moves, as suggested in
Ref.~\onlinecite{Scalettar:erg}, were used to assure ergodicity.  To
invert Eqs.~(\ref{eq:fug}) and (\ref{eq:ox}), the right-hand sides are
measured at the set of chemical potentials,
\begin{equation}
\mu(i) = \mu_\alpha \;+\; i\, \delta\mu,
\end{equation}
where $\mu_\alpha$ is the chemical potential used in the simulation,
$i=-7\ldots0\ldots7$ and $\delta\mu=0.02$.  After inversion and
averaging over configurations, particle sectors where $Z_N e^{\beta
\mu_\alpha N} / Z(\mu_\alpha) < 10^{-4}$ are dropped to prevent the
spread of roundoff error from the inversion.  The jackknife method was
used for error analysis.  It should be noted that, after analysis,
results at different values of $\mu$ are not statistically
independent.


\begin{figure}
\centerline{\includegraphics[width=3.5in]{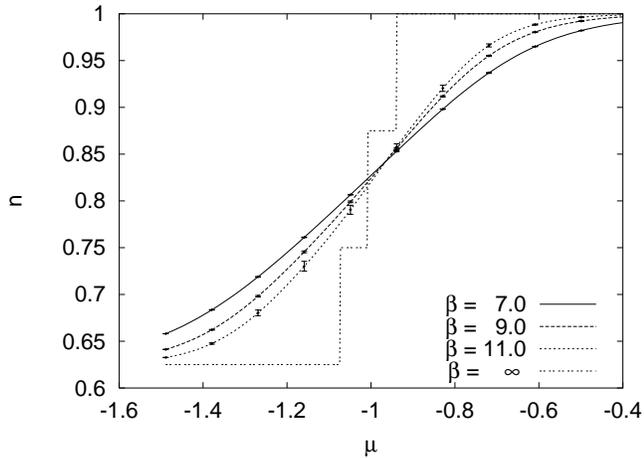}}
\caption{\label{fig:n} Density of the system for several different
  values of  $\beta$.  As errors depend on $\mu$, errorbars are shown
  at several points along the curves here and in subsequent figures.
Also shown is the zero-temperature result, calculated using exact
diagonalization.}
\end{figure}
\begin{figure}
\centerline{\includegraphics[width=3.5in]{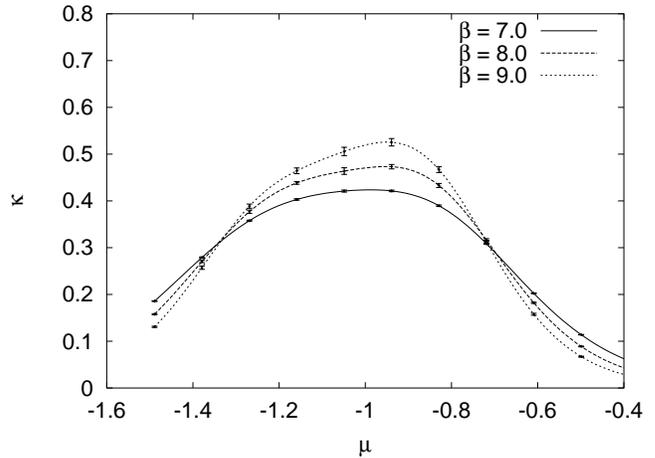}}
\caption{\label{fig:comp} Charge compressibility of the system,
  obtained by the analytical differentiation of $n$. 
}
\end{figure}

\begin{figure}
\centerline{\includegraphics[width=3.5in]{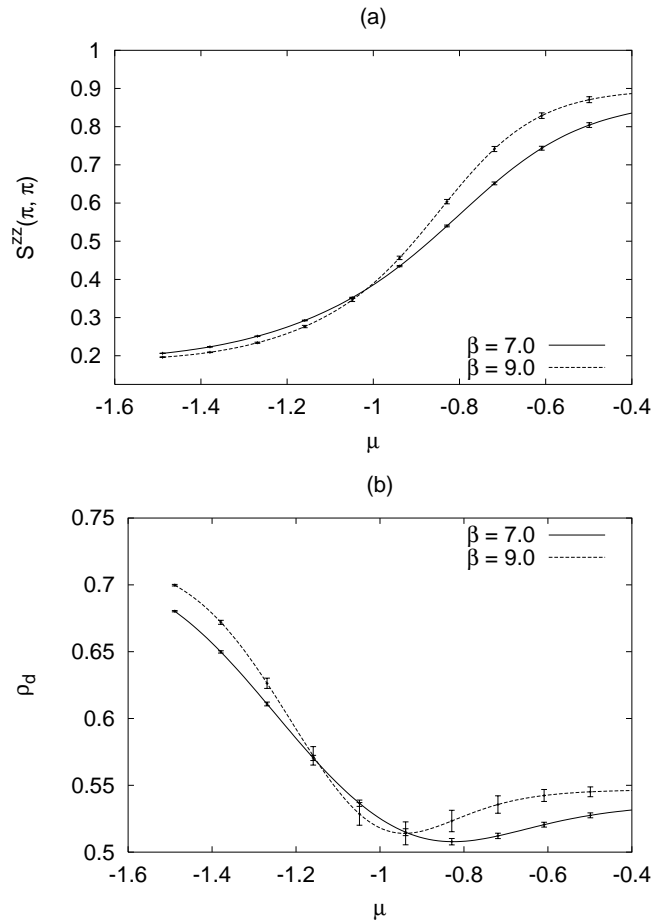}}
\caption{\label{fig:safpd} (a) 
 The antiferromagnetic structure factor, $S^{zz}(\pi, \pi)$. 
(b) The $d$-wave pair field correlation function.  
}
\end{figure}

In Fig.~\ref{fig:F} we plot the free energy difference, 
\begin{equation}
F(\mu)-F(0) =-\frac{1}{ \beta}\; \ln{\{Z(\mu)/Z(0)\}},
\end{equation}
as a function of $\mu$, for two values of $\beta$. In Fig.~\ref{fig:n} we
plot the density defined by $n(\mu) = \langle N \rangle/V$.
Here $V$ is the number of spatial lattice points and $\langle N
\rangle$ is calculated using the standard thermodynamic identity,  
\begin{equation}
  \langle N \rangle = -\frac{\partial F}{\partial \mu} =
  \frac{\sum_N N\; Z_N\; e^{\beta \mu N}}{\sum_N Z_N\; e^{\beta \mu N}}.
\label{eq:N}
\end{equation}
As the temperature is lowered, the transition between the half-filled
state ($n=1.0$) and the 6-hole state ($n=0.625$) becomes sharper.  In
particular, at zero-temperature the density decreases in a series of
jumps, due to the discreteness of the finite-size spectrum.


Within our framework it is also straightforward to compute the
compressibility of the system, $\kappa = \partial n / \partial \mu $,
by differentiation of Eq.~(\ref{eq:N}).  Note that 
the differentiation can be performed analytically.
Figure~\ref{fig:comp} shows the compressibility as a function of $\mu$
for different values of $\beta$.  As the temperature is lowered, the
compressibility develops a peak around $\mu=-1.0$ that is likely to be
the signature of the low-temperature divergence expected from the
metal-insulator transition.\cite{IMADA}



\begin{figure}
\centerline{\includegraphics[width=3.5in]{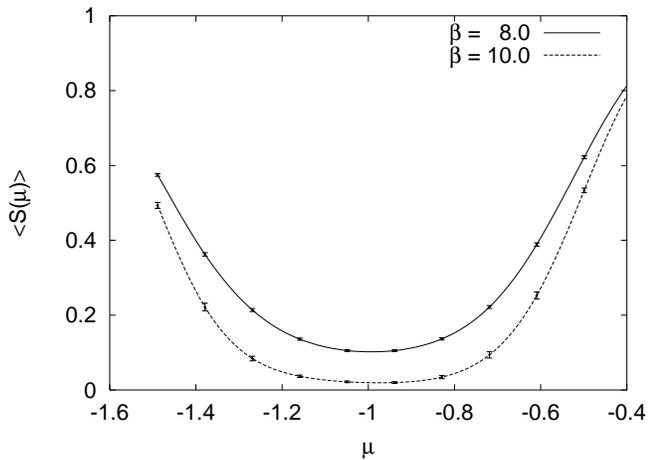}}
\caption{\label{fig:sign} The expectation value of the sign.
Calculated from runs with 100,000 Monte Carlo sweeps at $\mu =-1.5$,
$\mu =-1.025$, and $\mu=-0.6$.  }
\end{figure}



\enlargethispage{\baselineskip}
\looseness=-1
As previously mentioned, within our numerical scheme it is possible to
calculate observables that are not diagonal in the particle number.
Figure~\ref{fig:safpd}(a) shows the antiferromagnetic structure
factor.  This is given by
\begin{equation}
S^{zz}(\pi, \pi) =\frac{1}{V} \sum_{ij} (-1)^{i+j}\; S_i^z S_j^z,
\end{equation}
where $S_i^z = \frac{1}{2} c_{i\alpha}^\dagger \; \sigma_{\alpha
\beta}^z\; c_{i\beta}$ is the standard spin operator.  The plot of
this quantity vs.~$\mu$ clearly indicates that the antiferromagnetic
correlations present at half-filling are sharply suppressed upon
doping.  A similar plot of the equal-time $d$-wave pair field
correlation function is shown in Fig.~\ref{fig:safpd}(b).  Here the
$d$-wave pair field correlation function is given by
\begin{equation}
  \rho_d = \frac{1}{V} \sum_{ij} \Delta_i \Delta_j^\dagger
  \label{eq:pd}
\end{equation}
where $\Delta_i^\dagger=\frac{1}{2} \sum_\delta (-1)^\delta
c_{i\uparrow}^\dagger c_{i+\delta \downarrow}^\dagger $ creates two
electrons in a $d$-state.
Here, $\delta$ sums over the four near-neighbor sites of
$l$ and $(-1)^\delta$ gives the sign alternation characteristic of a
$d$-wave pairing amplitude.  The enhancement of $\rho_d$ toward
$\mu=0.0$ is a finite-size effect due to a strong antiferromagnetic
response in the nearest-neighbor terms in Eq.~(\ref{eq:pd}).

Finally, we show the expectation value of the sign in
Fig.~\ref{fig:sign}.  Here the sign is calculated as a continuous
function of $\mu$ using Eqs.~(\ref{eq:zp}) and (\ref{eq:zNp}).  Note
that the sign is small in the $\mu=-1.0$ region where the density is
changing rapidly and electron correlations are believed to be
important. The statistical fluctuations of the other observables grow
as the sign decreases.

\vspace{-0.2in}
\section{Conclusion}

In this paper we have presented a new method for extracting canonical
ensemble results from grand canonical ensemble quantum Monte Carlo
simulations.  As canonical information is only extracted from sectors
whose particle number is close to the average number of particles in the
 simulation, simulations must be performed at several
different chemical potentials to obtain results for a range of
particle number sectors.  These separate simulations can then be
combined to obtain a complete picture of the different canonical
ensembles with lower statistical fluctuations than any of the
simulations taken individually.  Once the canonical results are
obtained, they can be combined to give grand canonical results as a
continuous function of the chemical potential.

In this work we have presented results for the
two-dimensional Hubbard model on a $4\times 4$ lattice with
Coloumb interactions of moderate strength, but the method is
applicable to any quantum mechanical problem, simulated in the grand
canonical ensemble, for which particle number is conserved.  
 
\acknowledgements{
We would like to thank R.T.~Scalettar and A.~Sandvik
for insightful discussions and Federico Becca for help with the
Lanczos calculations.  This work was supported by the Department of
Energy under Grant \#DOE85-45197.}



\begin{thebibliography}{99}

\bibitem{REVIEW} E.W.~Carlson, V.J.~Emergy, S.A.~Kivelson, and D. Orgad,
{\it Concepts in High Temperature Superconductivity}, cond-mat/0206217.

\bibitem{DET} R.~Blankenbecler, D.J.~Sca\-la\-pi\-no and R.L.~Sugar,  
Phys.\ Rev.\ D
{\bf 24}, 2278, (1981);  and S.R.~White, D.J.~Sca\-la\-pi\-no, R.L.~Sugar, 
E.Y.~Loh, J.E.~Gubernatis and R.T.~Scalettar,  Phys.\ Rev.\ B {\bf 40}, 
506, (1989). 

\bibitem{FS}  A.M.~Ferrenberg and R.H.~Swendsen,  Phys.\ Rev.\ Lett. {\bf 63}, 
1195, (1989).

\bibitem{HIRSCH} J.E.~Hirsch, Phys. Rev. B {\bf 31}, 4403 (1985).

\bibitem{Scalettar:erg} R.T.~Scalettar, R.M.~Noack and R.P.~Singh,
   Phys.~Rev.~B {\bf 44}, 10502, (1991).

\bibitem{IMADA} N.~Furukawa and M.~Imada,
J.\ Phys.\ Soc.\ Jpn. {\bf 60}, 3604 (1991).



\end{thebibliography}
\end{document}